\documentclass[aps, reprint, superscriptaddress, amsmath,amssymb,showkeys,physrev]{revtex4-2}

\usepackage{array}
\usepackage{tabularx}
\usepackage{makecell}
\usepackage{booktabs}
\usepackage{dcolumn}
\usepackage{bm}
\usepackage{ragged2e}
\usepackage{graphicx}
\usepackage{xcolor}
\usepackage{colortbl}
\usepackage{xurl}
\usepackage[hidelinks]{hyperref}

\begin{document}

\title{\textbf{Optimizing Quantum Data Embeddings\\for Ligand-Based Virtual Screening}}

\author{Junggu Choi}
\email{choij14@ccf.org}
\affiliation{Department of Microbial Sciences in Health, Global Center for Pathogen and Human Health Research, Cleveland Clinic Research, Cleveland Clinic, OH, USA}

\author{Tak Hur}
\affiliation{Department of Statistics and Data Science, Yonsei University, Seoul, Republic of Korea}

\author{Seokhoon Jeong}
\affiliation{Department of Statistics and Data Science, Yonsei University, Seoul, Republic of Korea}

\author{Kyle L. Jung}
\affiliation{Department of Microbial Sciences in Health, Global Center for Pathogen and Human Health Research, Cleveland Clinic Research, Cleveland Clinic, OH, USA}

\author{Jun Bae Park}
\affiliation{Department of Microbial Sciences in Health, Global Center for Pathogen and Human Health Research, Cleveland Clinic Research, Cleveland Clinic, OH, USA}

\author{Junho Lee}
\affiliation{Department of Microbial Sciences in Health, Global Center for Pathogen and Human Health Research, Cleveland Clinic Research, Cleveland Clinic, OH, USA}

\author{Jae U. Jung}
\email{jungj14@ccf.org}
\affiliation{Department of Microbial Sciences in Health, Global Center for Pathogen and Human Health Research, Cleveland Clinic Research, Cleveland Clinic, OH, USA}

\author{Daniel K. Park}
\email{dkd.park@yonsei.ac.kr}
\affiliation{Department of Statistics and Data Science, Yonsei University, Seoul, Republic of Korea}
\affiliation{Department of Applied Statistics, Yonsei University, Seoul, Republic of Korea}
\affiliation{Department of Quantum Information, Yonsei University, Seoul, Republic of Korea}

\begin{abstract}
Effective molecular representations are essential for ligand-based virtual screening. We investigate how quantum data embedding strategies can improve this task by developing and evaluating a family of quantum–classical hybrid embedding approaches. These approaches combine classical neural networks with parameterized quantum circuits in different ways to generate expressive molecular representations and are assessed across two benchmark datasets of different sizes: the LIT-PCBA and COVID-19 collections. Across multiple biological targets and class-imbalance settings, several quantum and hybrid embedding variants consistently outperform classical baselines, especially in limited-data regimes. These results highlight the potential of optimized quantum data embeddings as data-efficient tools for ligand-based virtual screening.
\end{abstract}


\maketitle

\section{\label{sec:level1}Introduction}
Ligand-based virtual screening (LBVS) plays a central role in early-stage drug discovery, where millions of candidate molecules must be evaluated for potential biological activity \cite{ripphausen2011state, stahura2005new}. Because experimental screening is costly and slow, LBVS relies on computational models capable of capturing subtle structure–activity relationships under diverse conditions. Modern machine learning (ML) techniques have shown strong potential for improving LBVS based on their remarkable predictive capabilities \cite{baskin2020power, lavecchia2015machine, bahi2018deep}. For large-scale ligand library screening, ML models provide a significantly faster and more cost-efficient alternative to traditional biochemical validation processes \cite{carpenter2018deep}. Moreover, in terms of prediction accuracy for quantitative structure–activity relationship (QSAR) modeling or docking score estimation, ML approaches have outperformed conventional methods \cite{cieslak2024machine, bharath2025machine}. Despite these advantages, ML-driven drug discovery still faces several limitations. Many ML models are designed to learn from millions of ligand compounds, resulting in substantial computational demands. Yet, datasets collected from biological research rarely offer such scale; rather, they contain few labeled samples and are characterized by severe class imbalance, with positive compounds greatly outnumbered by negatives.

With recent advances in quantum computing, quantum machine learning (QML) has emerged as a promising paradigm that leverages quantum information processing to enhance data-driven learning beyond classical approaches \cite{biamonte2017quantum, schuld2015introduction, liu2021rigorous, huang2022quantum, huang2025generative}. By embedding data into high-dimensional quantum Hilbert spaces, QML models have the potential to represent and manipulate complex data distributions that may be difficult to capture with classical methods. In particular, quantum processes can sample from certain structured distributions more efficiently than known classical algorithms \cite{harrow2017quantum, aaronson2011computational}, suggesting potential advantages for learning tasks involving intricate or highly nonlinear patterns---properties that frequently arise in biological datasets. Moreover, several empirical studies have shown that QML models can exhibit improved performance under severe class imbalance, for example in anomaly detection \cite{oh2024quantum, kwon2025leveraging, carlsson2024enhancing}. These observations motivate the exploration of QML approaches for challenging data regimes commonly encountered in drug discovery.

When the target data are classical, quantum embedding constitutes the initial step, transforming classical inputs into quantum states to be processed by quantum algorithms \cite{lloyd2020quantum, havlivcek2019supervised}. Recent studies have highlighted that the quantum embedding step strongly affects the expressibility, training accuracy, robustness to noise, and generalization capacity of QML models \cite{schuld2021effect, caro2021encoding, hurunderstanding}. For this reason, the choice and optimization of the embedding strategy are of critical importance, as the resulting quantum states directly influence the subsequent computation and overall model performance. Given its central role, optimizing or appropriately selecting the quantum embedding method is essential to fully exploit the advantages of quantum computing for biological data analysis.
 
Among various approaches to quantum data embedding, Hur et al. \cite{hur2024neural} introduced a trainable method known as Neural Quantum Embedding (NQE), which leverages a classical neural network optimized through the principle of kernel–target alignment \cite{cristianini2001kernel}. In binary classification, the trace distance between quantum states from different classes is a key determinant of performance. NQE enhances inter-class distinguishability in a manner that cannot be achieved by completely positive and trace-preserving (CPTP) channels acting on fixed embeddings alone, guiding encoded states toward an ideal regime where trace distances approach 0 for same-class samples and 1 for different-class samples. This learned embedding strategy improves both training accuracy and generalization \cite{hurunderstanding} compared with fixed quantum feature maps by explicitly optimizing the geometry of the quantum representation used by downstream QML algorithms. The effectiveness of NQE has been demonstrated across a range of learning tasks in recent studies \cite{liu2025neural, kim2025multi, astuti2025use}.

Building on the strengths of trainable quantum embedding methods, we develop and evaluate a set of optimized quantum data embedding strategies for ligand-based virtual screening using two publicly available datasets: LIT-PCBA and COVID-19. Because these datasets differ substantially in size, we adopt separate evaluation schemes for each. For the larger LIT-PCBA dataset, we consider three categories of embeddings: (i) NQE implemented with ZZ and XYZ quantum feature map ansatzes; (ii) classical kernel-based embeddings, using a neural network–parameterized radial basis function (RBF) kernel; and (iii) quantum-pretrained classical embeddings, which reuse the neural networks obtained from NQE and adapt them within fully classical models through fine tuning or transfer-learning strategies. Each embedding method is followed by a binary classifier to assess representational quality: a quantum convolutional neural network (QCNN) for NQE-based embeddings, and a linear classifier for the classical and quantum-pretrained embeddings, chosen to maintain comparable model capacity. For the much smaller COVID-19 dataset, the limited sample size makes training NQE unlikely to be effective. Instead, we construct quantum kernels using parameterized embedding circuits and train classical support vector machines on these quantum kernels alongside classical RBF and linear kernels for comparison.

The remainder of this paper is organized as follows. Section~\ref{sec:level1} introduces the research scheme, including descriptions of the two datasets, the data embedding methods, and the binary classifier used to evaluate the embeddings and their classification performance. Section~\ref{sec:level2} presents the experimental results, including the classification performance of the algorithms. Section~\ref{sec:level3} discusses the performance of the binary classifiers and compares the applied embedding methods for each dataset. Finally, the conclusions, strengths, and limitations of this study are presented in Section~\ref{sec:level4}.

\section{\label{sec:level1}Methods}

\subsection{The overview of this research}
\begin{figure*}[htbp]
  \centering
  \includegraphics[width=\textwidth]{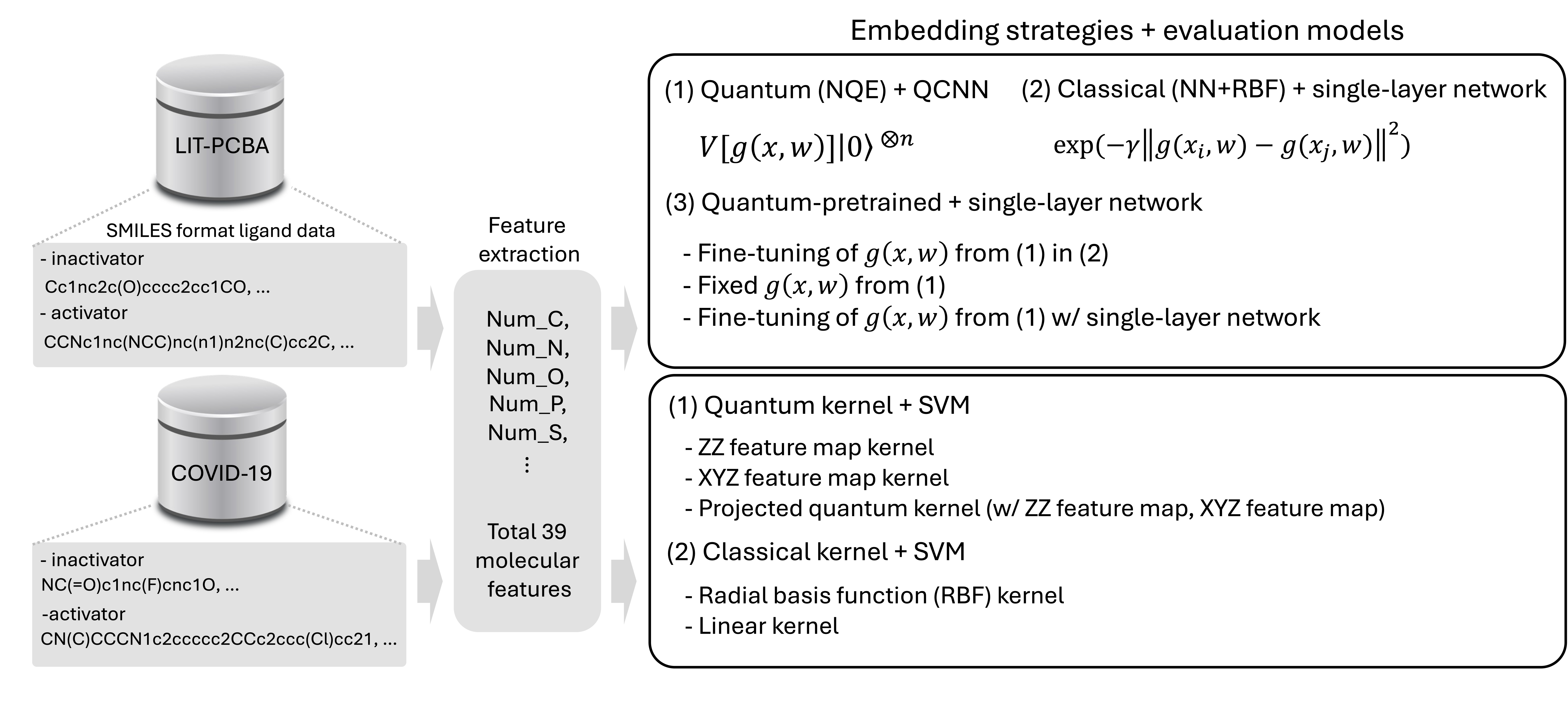}
  \caption{The overview of this research. $g(x,w)$ represent the classical neural network.}
  \label{fig:figure1}
\end{figure*}

In this study, we designed a research workflow consisting of three sequential steps. First, we collected two open-source datasets (LIT-PCBA and COVID-19) containing ligand information for biological targets, including both activators and inactivators \cite{tran2020lit, gawriljuk2021machine}. A common set of 39 molecular features, including the number of atoms and various physicochemical indices, was extracted from each dataset. Second, we implemented three embedding schemes (quantum, classical, and hybrid) based on the classical neural network trained with 39 molecular features from the LIT-PCBA dataset. Finally, a binary classifier was appended to the trained embeddings from the three conditions, and the classification performance was evaluated. In the same manner, the 39 features calculated from the COVID-19 dataset were evaluated by the SVM using the ZZ feature map and the XYZ feature map kernel, in comparison with the SVM including the RBF and linear kernel. The overview of this research is depicted in Figure~\ref{fig:figure1}.

\subsection{Dataset}
To examine the influence of different embedding approaches on LBVS, we selected two publicly available biological datasets: LIT-PCBA and COVID-19. The LIT-PCBA dataset contains ligand information in the Simplified Molecular Input Line Entry Specification (SMILES) format across 15 biological targets. The biological targets and the number of ligands in the LIT-PCBA dataset are summarized in Table~\ref{tab:table1}. Similarly, the COVID-19 dataset includes both 34 activators and 89 inhibitors, also represented in the SMILES format.

\begin{table*}[htbp]
\renewcommand{\arraystretch}{1.5}
\caption{\label{tab:table1} Biological targets in the LIT-PCBA dataset}
\begin{ruledtabular}
\begin{tabular}{lccc}
\textbf{No.} & 
\makecell{\textbf{Target}\\\textbf{(\# of activators / \# of inactivators)}} & 
\textbf{No.} & 
\makecell{\textbf{Target}\\\textbf{(\# of activators / \# of inactivators)}} \\
\colrule
1 & MAPK1 (308/62,629) & 9 & PKM2 (546/245,523) \\
2 & TP53 (79/4,168) & 10 & VDR (884/355,388) \\
3 & FEN1 (369/355,402)& 11 & ALDH1 (7,168/137,965) \\
4 & ESR1 antagonist (102/4,948) & 12 & IDH1 (39/362,409) \\
5 & ADRB2 (17/312,483) & 13 & KAT2A (194/348,548) \\
6 & GBA (166/296,502) & 14 & PPARG (27/5,211) \\
7 & MTORC1 (97/32,972) & 15 & ESR1 agonist (13/5,583) \\
8 & OPRK1 (24/269,816)&  &  \\
\end{tabular}
\end{ruledtabular}
\end{table*}

\subsection{Data preprocessing and the feature extraction}
Among the 15 biological targets in the LIT-PCBA dataset, we selected 8 targets that contain more than 100 activator ligands: GBA, KAT2A, ESR1 antagonist, MAPK1, FEN1, PKM2, VDR, and ALDH1. From the SMILES representations of ligands, 39 molecular descriptors were computed as features, following the approaches widely adopted in previous studies \cite{lo2018machine, todeschini2008handbook, bajorath2002integration, hong2008mold2, sawada2014benchmarking}. These molecular descriptors include various chemical and structural properties, including atomic composition, structural topology, physicochemical characteristics, basic electronic information, and molecular complexity. The complete list of the 39 molecular descriptors is summarized in~\ref{appendix:appendixA}. For the COVID-19 dataset, the same set of features was calculated to ensure consistency in the comparative analysis.

\subsection{Data embedding methods and binary classifiers}
\subsubsection{Neural quantum embedding and the quantum convolutional neural network}
Among the various quantum embedding methods introduced in previous studies, we selected NQE to evaluate the embedding capacity for the molecular features calculated from the activator and inactivator. This embedding method has 2 major advantages. First, unlike conventional fixed embedding circuits, this approach leverages a classical neural network to construct a trainable and flexible embedding framework, enabling its application to a wide range of classical datasets. Second, to enhance the baseline performance of subsequent QML algorithms, the embedding circuit is trained jointly with the classical neural network to maximize the trace distance, which represents the distinguishability between two quantum states. For training NQE, a quantum fidelity-based loss function is designed as follows:

\begin{eqnarray}
w' &=& \arg \min_{w}
\sum_{ij}
\left[|\langle x_i(w) | x_j(w) \rangle|^2 - \frac{1}{2}(1 + y_i y_j) \right]^2
\label{eq:formula1}
\end{eqnarray}

In Eq.~(\ref{eq:formula1}), $w'$ and $w$ denote the trainable parameters of the classical neural network, and $|x_j(w)\rangle$ represents the quantum state that encodes the classical data through the neural network and the embedding circuit ($|x_j(w)\rangle = V[g(x_j,w)]|0\rangle^{\otimes n}$). $y_i$ and $y_j$ indicate the class labels of the classical data samples $x_i$ and $x_j$, respectively. In this study, we employed the ZZ feature map and the XYZ feature map, which are known to be classically intractable, for the NQE implementation \cite{liu2025neural, havlivcek2019supervised}. The ZZ feature map has been widely adopted in QML-based drug discovery applications \cite{mensa2023quantum, batra2021quantum}, and the XYZ feature map demonstrated its effectiveness in binary classification tasks combined with NQE in a previous study \cite{liu2025neural}. A linear structure composed of 2-qubit gates was used for both embedding circuits to check the applicability of the most simple structure. The corresponding unitary operators of these embedding circuits are defined as

\begin{figure*}[htbp]
  \centering
  \includegraphics[width=\textwidth]{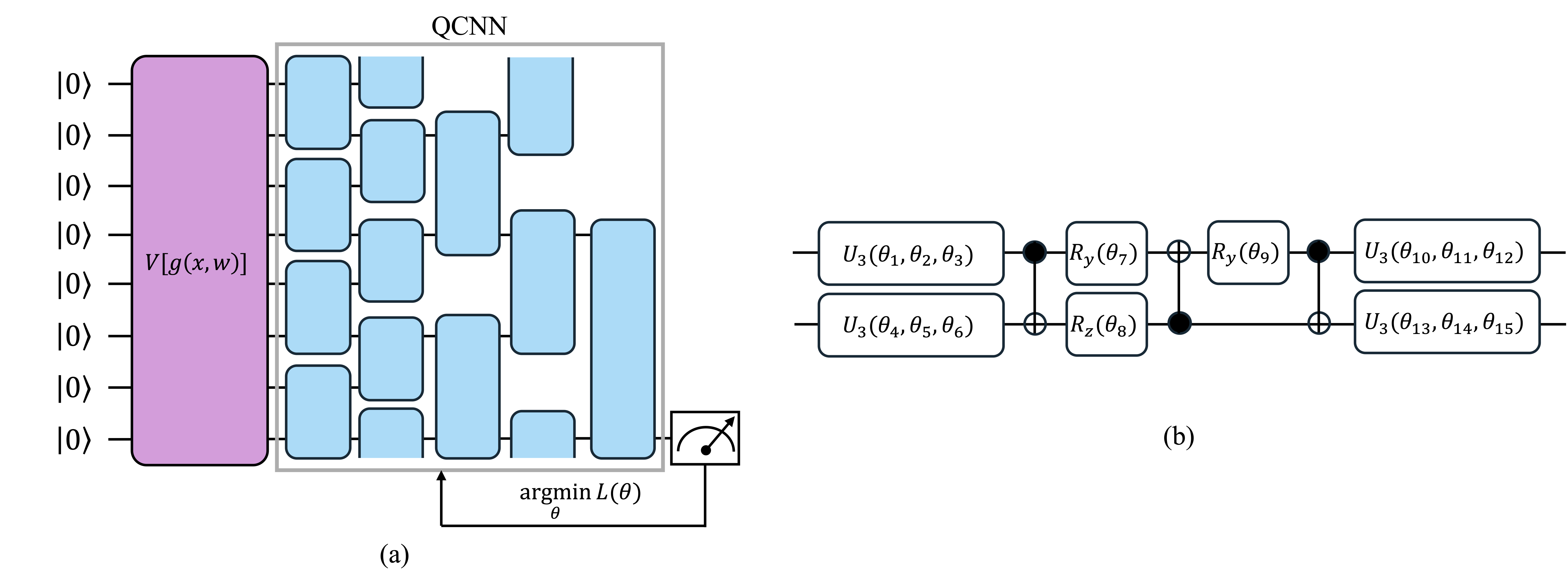}
  \caption{(a) Schematic representation of the 8-qubit QCNN integrated with neural quantum embedding (NQE). The purple section indicates the NQE, which transforms the classical data $x$ into the quantum state $|x\rangle$. The blue section represents the architecture of the 8-qubit quantum convolutional neural network (QCNN). (b) Structural diagram of SU(4) ansatz. $U_3(\theta_1,\theta_2,\theta_3)$ denotes an arbitrary single-qubit gate. It can be expressed by $U_3(\theta_1,\theta_2,\theta_3)=R_z(\theta_2)R_x(-\pi/2)R_z(\theta_1)R_x(\pi/2)R_z(\theta_3)$. $R_i(\theta)$ represents the rotation by $\theta$ around the $i$-axis of the Bloch sphere.}
  \label{fig:figure2}
\end{figure*}

\begin{widetext}
\begin{equation}
V_{ZZ}(\phi(x))
=
\Big[
\exp\Big(
 i \sum_i \phi_i(x) Z_i
 + i \sum_{i,j} \phi_{i,j}(x) Z_i Z_j
\Big)
 H^{\otimes n}
\Big]^l
\label{eq:formula2}
\end{equation}

\begin{equation}
\begin{aligned}
V_{XYZ}(\phi(x))
&=
\Bigg\{
\exp\Big[
 i \sum_k \phi_k(x) Z_k
 + \phi_{n+k}(x) Z_k Z_{k+1}
\Big] \\
&\quad
\exp\Big[
 i \sum_k \phi_k(x) Y_k
 + \phi_{n+k}(x) Y_k Y_{k+1}
\Big] \\
&\quad
\exp\Big[
 i \sum_k \phi_k(x) X_k
 + \phi_{n+k}(x) X_k X_{k+1}
\Big]
\Bigg\}^l
\end{aligned}
\label{eq:formula3}
\end{equation}
\end{widetext}

\noindent
where $\phi_i(x)=x_i$ and $\phi_{i,j}(x)=(\pi-x_i)(\pi-x_j)/2$. In this work, we adopted these commonly used functions for $\phi_i(x)$ and $\phi_{i,j}(x)$. In addition, we chose $l$=3 for the ZZ feature map and $l$=2 for the XYZ feature map. After the classical data embedding by NQE, we evaluated the binary classification performance of the 8-qubit QCNN. For the convolutional layer of the QCNN, we employed an SU(4) ansatz capable of representing any arbitrary two-qubit unitary operation. In general, a unitary in the SU(4) group can be decomposed into at most three CNOT gates together with 15 elementary single-qubit rotations \cite{vatan2004optimal, maccormack2022branching}. The schematic representation of the 8-qubit QCNN with NQE and SU(4) ansatz structure is depicted in Figure~\ref{fig:figure2}.

\subsubsection{Neural network-parameterized RBF kernel}
As a classical counterpart of the QCNN with NQE, we combined the classical neural network with the RBF kernel to make it trainable \cite{wilson2016deep}. This approach was developed based on the concept of kernel–target alignment, where the kernel function is optimized to increase the similarity between kernel-induced feature space and label distribution. The overall procedure of the trainable RBF kernel can be summarized as follows: 

Let $x_i \in \mathbb{R}^{39}$ denote the input feature vector consisting of 39 molecular descriptors calculated from each ligand sample, and let $f_{\mathrm{NN}}(x_i; w)$ represent the classical neural network with trainable parameters $w$. The output of the neural network can be expressed as $h_i = f_{\mathrm{NN}}(x_i; w)$. $h_i \in \mathbb{R}^p$ denotes the output feature vector from the classical neural network. Using the output features $\{h_i\}_{i=1}^n$, the RBF kernel matrix is defined as

\begin{equation}
    K_{ij} = \exp(-\gamma \|h_i - h_j\|^2)
    \label{eq:rbf_kernel}
\end{equation}
where $\gamma$ is the kernel width hyperparameter. For the simplicity, we fixed the kernel parameter $\gamma=1$. 

To align the kernel matrix with the label similarity, the loss function is defined as

\begin{equation}
    \mathcal{L} = \frac{1}{M}\sum_{i<j}\left[K_{ij} - \delta_{y_iy_j}\right]^2
    \label{eq:rbf_loss}
\end{equation}

where $y_i \in \{-1, +1\}$ denotes the class label, and $M$ represents the number of unique elements in the upper triangular part of $K$. Minimizing this loss enables the neural network to learn a representation that maximizes the separability of data samples in the kernel-induced space while preserving pairwise class relationships. To evaluate the embedding efficiency of the trainable RBF kernel, we assessed the classification performance of the trained classical neural network using a single-layer network.

\subsubsection{Quantum-pretrained classical embeddings}
\begin{figure*}[htbp]
  \centering
  \includegraphics[width=0.98\textwidth]{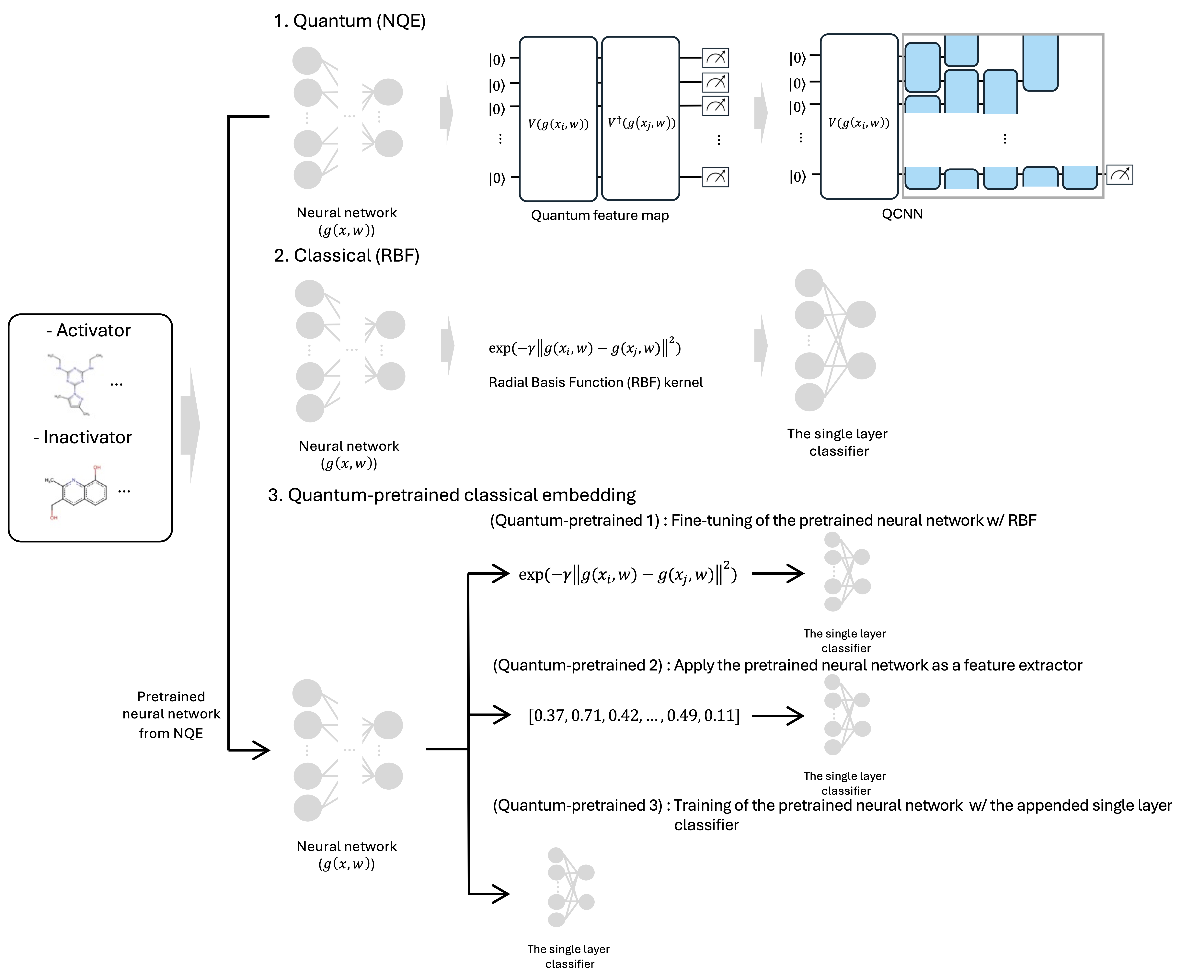}
  \caption{The diagram of the data embedding methods applied to the LIT-PCBA dataset.}
  \label{fig:figure3}
\end{figure*}

Based on the two previously introduced directions (NQE and the neural network-parameterized RBF kernel), we developed three different hybrid approaches. First, the trained classical neural network from NQE was fine-tuned using the neural network-parameterized RBF kernel. In this setting, the pretrained weights obtained from NQE were further updated through the trainable RBF-based optimization. After fine-tuning, a single-layer network was appended to the fine-tuned neural network to evaluate its classification performance. Second, a single-layer network was appended to the pretrained neural network from NQE, while keeping the NQE-trained weights fixed. Only the parameters of the single-layer network were trained to assess the classification performance. Third, similar to the second approach, a single-layer network was added to the pretrained neural network from NQE. Unlike the second approach, both the NQE-trained neural network and the single-layer network were jointly trained in this case. All embedding conditions for the LIT-PCBA dataset using NQE, the trainable RBF, and the hybrid conditions are described in Table~\ref{tab:table3} and Figure~\ref{fig:figure3}.

\begin{table*}[htbp]
\renewcommand{\arraystretch}{1.5}
\caption{\label{tab:table3} The list of embedding methods evaluated on the LIT-PCBA dataset.}
\begin{ruledtabular}
\begin{tabular}{l l l}
\textbf{Category} & \textbf{Embedding method} & \textbf{Binary classifier} \\
\colrule
Quantum & NQE\footnotemark[1] (the ZZ feature map or the XYZ feature map) & QCNN\footnotemark[2] \\
Classical & Trainable RBF\footnotemark[3] kernel with the classical NN\footnotemark[4] & Single-layer NN \\
Quantum-pretrained 1 & Fine-tuning of the NQE-pretrained NN using the trainable RBF kernel & Single-layer NN \\
Quantum-pretrained 2 & NQE-pretrained NN with fixed weights & Single-layer NN \\
Quantum-pretrained 3 & Joint training of the NQE-pretrained NN and appended single-layer & Single-layer NN \\
\end{tabular}
\end{ruledtabular}
\footnotetext[1]{NQE: the neural quantum embedding}
\footnotetext[2]{QCNN: the quantum convolutional neural network}
\footnotetext[3]{RBF: the radial basis function}
\footnotetext[4]{NN: the neural network}
\end{table*}

\subsubsection{Quantum kernels using the ZZ feature map and the XYZ feature map} 
Unlike the LIT-PCBA analysis---which evaluated NQE, RBF, and quantum-pretrained embedding schemes---the COVID-19 dataset required a different approach due to its small size and the quadratic computational cost of constructing kernel matrices for large datasets. Accordingly, we analyzed the COVID-19 data using SVMs equipped with 4- and 8-qubit quantum kernels. Two types of quantum kernels were constructed: one derived from the ZZ feature map and the other from the XYZ feature map. The quantum kernel matrix was computed using the standard fidelity-based definition, where each entry corresponds to the overlap between two embedded quantum states, i.e., $K(x_i, x_j) = |\langle \psi(x_i) | \psi(x_j) \rangle|^2$.

In addition, to further evaluate the embedding efficiency, we additionally applied the projected quantum kernel (PQK) proposed by ~\cite{huang2021power}. In this approach, classical data are first encoded into quantum states through an embedding circuit. Unlike the conventional quantum kernel, which measures the global fidelity between two encoded states, PQK maps these states back into a reduced classical representation via measurements of local observables.  This projection enhances the geometric distinction between data points in the kernel-induced feature space while maintaining the quantum expressivity of the embedding. In our implementation, we followed the PQK formulation based on the one-particle reduced density matrix (1-RDM), where the kernel is defined as

\begin{eqnarray}
k_{\mathrm{PQ}}(x_i, x_j) = \exp\!\Big[-\gamma \sum_k \|\rho_k(x_i) - \rho_k(x_j)\|_F^2\Big],
\label{eq:pqk}
\end{eqnarray}

\noindent
with $\rho_k(x) = \mathrm{Tr}_{j\neq k}[\rho(x)]$ and the scaling parameter $\gamma$. We calculated the $\gamma$ value from the variance of all measured expectation values of the Pauli operators as follows:

\begin{eqnarray}
\gamma = \frac{1}{\mathrm{Var}(v) \cdot d}
\label{eq:gamma_pqk}
\end{eqnarray}
where $\mathrm{Var}(v)$ denotes the variance of the concatenated expectation values of the Pauli operators $v = \{\langle X_i \rangle, \langle Y_i \rangle, \langle Z_i \rangle\}_{i=1}^{q}$, and $d$ represents the number of input features. The embedding conditions with the QSVM for the COVID-19 dataset are listed in Table~\ref{tab:table4}.

\begin{table*}[htbp]
\renewcommand{\arraystretch}{1.5}
\caption{\label{tab:table4} The list of embedding methods evaluated on the COVID-19 dataset.}
\begin{ruledtabular}
\begin{tabular}{l l l}
\textbf{Category} & \textbf{Embedding method} & \textbf{Binary classifier} \\
\colrule
Quantum & ZZ feature map kernel & SVM\footnotemark[1] \\
Quantum & PQK\footnotemark[2] with the ZZ feature map & SVM \\
Quantum & XYZ feature map kernel & SVM \\
Quantum & PQK with the XYZ feature map & SVM \\
Classical & RBF\footnotemark[3] kernel & SVM \\
Classical & linear kernel & SVM \\
\end{tabular}
\end{ruledtabular}
\footnotetext[1]{SVM: the support vector machine}
\footnotetext[2]{PQK: the projected quantum kernel}
\footnotetext[3]{RBF: the radial basis function}
\end{table*}

\subsection{Model training and performance evaluation}
All experiments in this study were conducted through numerical simulations. For the LIT-PCBA dataset, to examine the influence of class imbalance between activators and inactivators, all experimental conditions were evaluated under two class ratio settings: a balanced 1:1 ratio and an imbalanced 1:6 ratio (activator:inactivator). In the case of the COVID-19 dataset, the analysis was performed using the original sample distribution without balancing, due to the relatively small number of available samples. For the NQE-based experiments on the LIT-PCBA dataset, the structure of the classical neural network was optimized with an input layer of 39 dimensions (corresponding to the molecular descriptors) and an output layer matching the number of classical features in the ZZ feature map and the XYZ feature map. In addition, the NQE training was stopped early when validation loss showed no improvement for 40 epochs to prevent overfitting. To evaluate the classification performance of the single-layer network as a binary classifier, the network was trained using the binary cross-entropy with logit loss function and Adam optimizer implemented in PyTorch. Model performance was evaluated as the mean $\pm$ standard deviation of accuracy and balanced accuracy obtained across five independent repetitions, where each repetition uses a distinct random initialization of the algorithm. All hyperparameters, including the learning rate, batch size, and number of iterations, were optimized using random search.

\subsection{Tools}
All implementations were conducted using Python~3.9. Molecular feature extraction and descriptor calculations were performed using RDKit~2024.03.6 (release date: November~7,~2024). Quantum and quantum-classical hybrid embedding models were implemented with PennyLane~0.42.0 and PyTorch~2.7.1. 

\section{\label{sec:level2}Results}
\subsection{Quantum and classical embedding results for the LIT-PCBA dataset}
\begin{figure*}[htbp]
  \centering
  \includegraphics[width=0.9\textwidth]{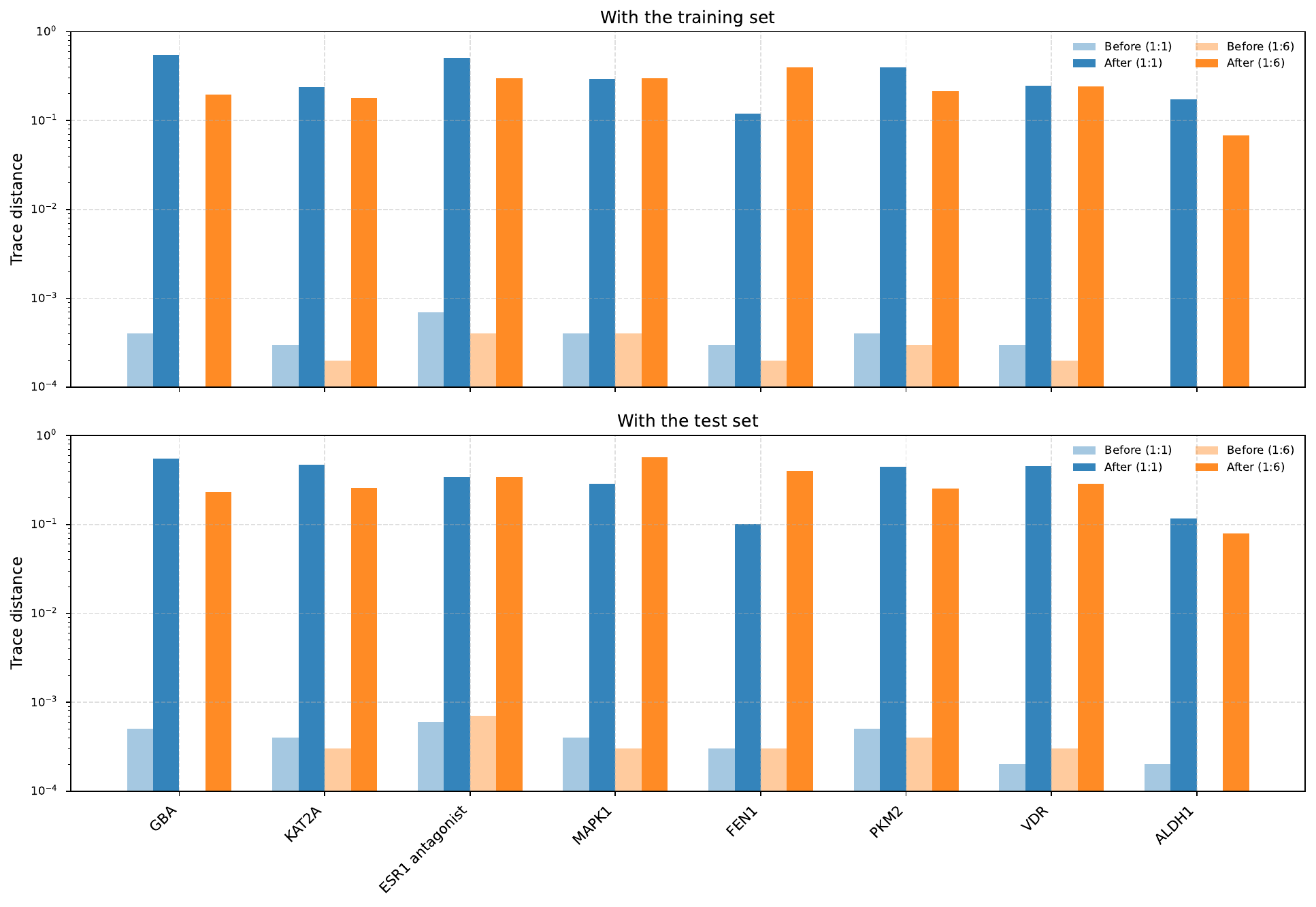}
  \caption{Trace distance changes before and after training of NQE with the ZZ feature map. “Before” and “After” indicate the trace distances of NQE prior to and following training, respectively. “1:1” and “1:6” denote the class ratios of the dataset.}
  \label{fig:figureZZ_TD}
\end{figure*}

\begin{figure*}[htbp]
  \centering
  \includegraphics[width=0.9\textwidth]{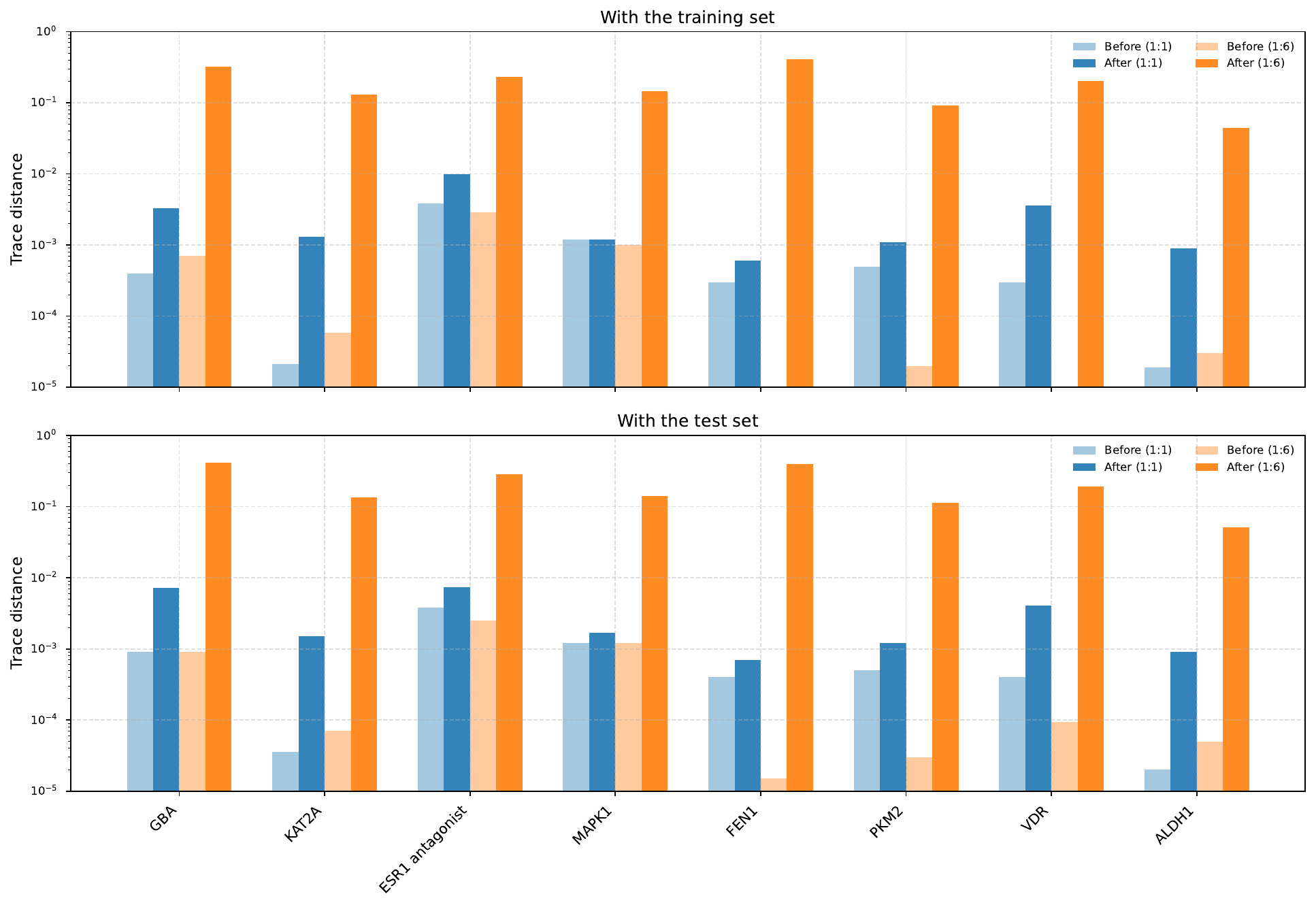}
  \caption{Trace distance changes before and after training of NQE with the XYZ feature map. “Before” and “After” indicate the trace distances of NQE prior to and following training, respectively. “1:1” and “1:6” denote the class ratios of the dataset.}
  \label{fig:figureXYZ_TD}
\end{figure*}

In order to evaluate the embedding performance of NQE, the classical neural network architecture combined with the ZZ feature map and the XYZ feature map was optimized by minimizing the fidelity-based loss function. The trace distances between quantum states were calculated before and after training to assess the optimization of NQE. For NQE with the ZZ and XYZ feature map, the trace distance values increased after training under all conditions. In the balanced class ratio (1:1) condition, the GBA target showed the largest increase, from 0.0004/0.0005 (trace distance of the train dataset / trace distance of the test dataset) to 0.5444/0.5535. The KAT2A and ESR1 antagonist targets also showed increased trace distances after training (KAT2A: 0.0003/0.0004 to 0.2376/0.4714; ESR1 antagonist: 0.0007/0.0006 to 0.5080/0.3435). Under the unbalanced class ratio (1:6) condition, MAPK1 and FEN1 showed the largest increases (MAPK1: 0.0004/0.0003 to 0.2992/0.5676; FEN1: 0.0002/0.0003 to 0.3972/0.4019), while other targets such as PKM2, VDR, and ALDH1 showed smaller changes. Changes in the trace distance obtained from the NQE with the ZZ feature map are shown in Figure~\ref{fig:figureZZ_TD}. In the case of NQE combined with the XYZ feature map, the increase in trace distance after training was smaller than that of the ZZ feature map. In the balanced class ratio condition, ESR1 antagonist increased from 0.0038/0.0038 to 0.0098/0.0074, while other targets such as GBA and KAT2A remained close to their initial values. Under the unbalanced condition, GBA and FEN1 showed higher post-training trace distances compared to other targets (GBA: 0.0007/0.0009 to 0.3199/0.4161; FEN1: 1.0e-05/1.5e-05 to 0.4132/0.3987). The overall results indicate that NQE effectively increased the separability of encoded quantum states, particularly with the ZZ feature map configuration. Increased trace distances from NQE with the XYZ feature map are presented in Figure~\ref{fig:figureXYZ_TD}. The detailed trace distance values with the ZZ and XYZ feature map for each target are summarized in~\ref{appendix:appendixA}. In addition, to confirm the convergence of NQE models, we checked trajectories of the training and validation loss value for each condition. Most of loss curves indicated stable convergence across all NQE configurations. For the comparison with the quantum embedding method, RBF kernel with the classical neural network was implemented using the same classical neural network architecture as in NQE. The training and validation loss curves showed an overall decreasing trend in the loss values during optimization. All training and validation loss curves are shown in~\ref{appendix:appendixC}. Applied hyperparameters and structures of the classical neural network used for optimizing both the quantum and classical embedding models are summarized in~\ref{appendix:appendixB}.

Following the optimized embedding method, we evaluated the classification performance of the QCNN with the pretrained NQE and the single-layer network with the the pretrained RBF kernel for each target. For the pretrained NQE with the ZZ feature map, QCNN showed overall higher classification accuracies than the single-layer network with the RBF kernel across most biological targets (Table~\ref{tab:table7}). Overall, the quantum model utilizing the pretrained NQE based on the ZZ feature map showed comparatively higher classification performance across most datasets. In the case of the pretrained NQE with the XYZ feature map, at the 1:1 ratio, the single-layer network model with the pretrained RBF kernel achieved higher accuracies for all targets (Table~\ref{tab:table8}). On the other hand, at the 1:6 ratio, QCNN showed higher values for GBA, ESR1 antagonist, MAPK1, FEN1, and VDR. All training loss curves of QCNN and the single-layer network are listed in~\ref{appendix:appendixC}.

\begin{table*}[htbp]
\caption{\label{tab:table7} Classification performances of the QCNN with the pretrained NQE (ZZ feature map) and the single-layer network with the pretrained RBF kernel. QCNN metrics outperforming the RBF baseline are highlighted in bold.}
\centering
\begin{tabularx}{\textwidth}{
l
c
c
c
c
}
\toprule
\textbf{Target} &
\makecell{\textbf{QCNN with NQE$^{\mathrm{a}}$} \\ \textbf{(1:1 ratio)}} &
\makecell{\textbf{Single-layer network} \\ \textbf{with RBF$^{\mathrm{b}}$} \\ \textbf{(1:1 ratio)}} &
\makecell{\textbf{QCNN with NQE} \\ \textbf{(1:6 ratio)}} &
\makecell{\textbf{Single-layer network} \\ \textbf{with RBF} \\ \textbf{(1:6 ratio)}}\\
\midrule
GBA & \textbf{0.81 $\pm$ 0.01} & 0.54 $\pm$ 0.84 & 0.60 $\pm$ 0.06 & \textbf{0.65 $\pm$ 0.02} \\
KAT2A & 0.52 $\pm$ 0.06 & \textbf{0.56 $\pm$ 0.01} & \textbf{0.63 $\pm$ 0.04} & 0.55 $\pm$ 0.02 \\ 
ESR1 antagonist & \textbf{0.75 $\pm$ 0.01} & 0.56 $\pm$ 0.04 & \textbf{0.74 $\pm$ 0.05} & 0.61 $\pm$ 0.04 \\
MAPK1 & \textbf{0.69 $\pm$ 0.01} & 0.57 $\pm$ 0.55 & 0.50 $\pm$ 0.02 & \textbf{0.58 $\pm$ 0.01} \\
FEN1 & \textbf{0.55 $\pm$ 0.01} & 0.53 $\pm$ 0.01 & \textbf{0.77 $\pm$ 0.02} & 0.60 $\pm$ 0.04 \\
PKM2 & \textbf{0.71 $\pm$ 0.00} & 0.57 $\pm$ 0.01 & \textbf{0.69 $\pm$ 0.02} & 0.50 $\pm$ 0.02 \\
VDR & 0.52 $\pm$ 0.01 & \textbf{0.58 $\pm$ 0.02} & \textbf{0.68 $\pm$ 0.01} & 0.57 $\pm$ 0.03 \\
ALDH1 & 0.56 $\pm$ 0.00 & \textbf{0.60 $\pm$ 0.00} & \textbf{0.55 $\pm$ 0.03} & 0.55 $\pm$ 0.02 \\
\bottomrule
\end{tabularx}
\vspace{2pt}
\raggedright
\scriptsize
$^{\mathrm{a}}$ NQE: pretrained neural quantum embedding with the ZZ feature map.  
$^{\mathrm{b}}$ RBF: pretrained radial basis function kernel with a classical neural network.
\end{table*}

\begin{table*}[htbp]
\caption{\label{tab:table8} Classification performances of the QCNN with the pretrained NQE (XYZ feature map) and the single-layer network with the pretrained RBF kernel. QCNN metrics outperforming the RBF baseline are highlighted in bold.}
\centering
\begin{tabularx}{\textwidth}{
l
c
c
c
c
}
\toprule
\textbf{Target} &
\makecell{\textbf{QCNN with NQE$^{\mathrm{a}}$} \\ \textbf{(1:1 ratio)}} &
\makecell{\textbf{Single-layer network} \\ \textbf{with RBF$^{\mathrm{b}}$} \\ \textbf{(1:1 ratio)}} &
\makecell{\textbf{QCNN with NQE} \\ \textbf{(1:6 ratio)}} &
\makecell{\textbf{Single-layer network} \\ \textbf{with RBF} \\ \textbf{(1:6 ratio)}}\\
\midrule
GBA & 0.50 $\pm$ 0.01 & \textbf{0.52 $\pm$ 0.15} & \textbf{0.80 $\pm$ 0.01} & 0.64 $\pm$ 0.01 \\
KAT2A & 0.52 $\pm$ 0.01 & \textbf{0.54 $\pm$ 0.04} & 0.54 $\pm$ 0.04 & \textbf{0.60 $\pm$ 0.01} \\
ESR1 antagonist & 0.49 $\pm$ 0.01 & \textbf{0.75 $\pm$ 0.06} & \textbf{0.69 $\pm$ 0.04} & 0.65 $\pm$ 0.01 \\
MAPK1 & 0.50 $\pm$ 0.00 & \textbf{0.57 $\pm$ 0.03} & \textbf{0.63 $\pm$ 0.07} & 0.54 $\pm$ 0.00 \\
FEN1 & 0.50 $\pm$ 0.00 & \textbf{0.63 $\pm$ 0.03} & \textbf{0.81 $\pm$ 0.05} & 0.55 $\pm$ 0.05 \\
PKM2 & 0.50 $\pm$ 0.00 & \textbf{0.61 $\pm$ 0.00} & 0.62 $\pm$ 0.06 & \textbf{0.66 $\pm$ 0.01} \\
VDR & 0.50 $\pm$ 0.00 & \textbf{0.60 $\pm$ 0.01} & \textbf{0.65 $\pm$ 0.01} & 0.58 $\pm$ 0.01 \\
ALDH1 & 0.50 $\pm$ 0.02 & \textbf{0.58 $\pm$ 0.01} & 0.56 $\pm$ 0.05 & \textbf{0.58 $\pm$ 0.01} \\
\bottomrule
\end{tabularx}
\vspace{2pt}
\raggedright
\scriptsize
$^{\mathrm{a}}$ NQE: pretrained neural quantum embedding with the ZZ feature map.  
$^{\mathrm{b}}$ RBF: pretrained radial basis function kernel with a classical neural network.
\end{table*}

\subsection{Quantum-pretrained classical embedding approaches based on trained neural networks}

From the trained neural networks in the quantum and classical embedding approaches, we further evaluated three hybrid embedding methods that combined the pretrained classical neural networks with the single-layer network. For the hybrid models utilizing the classical neural networks trained on the NQE with the ZZ feature map (Table~\ref{tab:table9}), the overall classification performance varied depending on the biological target and class ratio. Under the balanced (1:1) class ratio, quantum-derived 2 and quantum-derived 3 achieved higher accuracies for KAT2A, ESR1 antagonist, FEN1, and VDR compared to both the quantum and classical conditions (KAT2A: 0.69 $\pm$ 0.06, ESR1 antagonist: 0.80 $\pm$ 0.02, FEN1: 0.62 $\pm$ 0.01, and VDR: 0.67 $\pm$ 0.02). For the unbalanced (1:6) class ratio, Hybrid 3 outperformed the previous embedding approaches for GBA, KAT2A, and MAPK1 (GBA: 0.79 $\pm$ 0.01, KAT2A: 0.66 $\pm$ 0.02, and MAPK1: 0.67 $\pm$ 0.03). In addition, ESR1 antagonist achieved 0.80 $\pm$ 0.01 with quantum-derived 1, which was higher than the quantum and classical baselines. In quantum-derived 2, VDR and ALDH1 also showed improved performance compared to the quantum and classical results (VDR: 0.70 $\pm$ 0.03 and ALDH1: 0.59 $\pm$ 0.01). From the hybrid models trained on the NQE with the XYZ feature map (Table~\ref{tab:table10}), under the balanced (1:1) class ratio, GBA achieved the highest accuracy of 0.65 $\pm$ 0.01 with quantum-derived 1, KAT2A and PKM2 also reached 0.60 $\pm$ 0.00 and 0.63 $\pm$ 0.00 with quantum-derived 2 and quantum-derived 3, respectively. Under the unbalanced (1:6) class ratio, PKM2 and VDR achieved the highest accuracies of 0.73 $\pm$ 0.01 and 0.69 $\pm$ 0.00, respectively, both exceeding the baseline quantum and classical results. In addition, KAT2A and ESR1 antagonist exhibited relatively strong performances of 0.65 $\pm$ 0.02 and 0.75 $\pm$ 0.02 with quantum-derived 2. All hyperparameters for hybrid conditions and loss curves are listed in~\ref{appendix:appendixC}.

\begin{table*}[htbp]
\caption{\label{tab:table9}
Classification performances of quantum-pretrained classical embedding methods (trained classical neural networks from NQE with the ZZ feature map). Quantum-pretrained performance outperforming both NQE and RBF is highlighted in bold.}
\scriptsize
\centering
\begin{tabularx}{\textwidth}{lcccccccc}
\toprule
\textbf{Target} &
\textbf{Class ratio} &
\makecell{\textbf{Quantum-}\\\textbf{pretrained 1}} &
\makecell{\textbf{Quantum-}\\\textbf{pretrained 2}} &
\makecell{\textbf{Quantum-}\\\textbf{pretrained 3}} &
\textbf{Class ratio} &
\makecell{\textbf{Quantum-}\\\textbf{pretrained 1}} &
\makecell{\textbf{Quantum-}\\\textbf{pretrained 2}} &
\makecell{\textbf{Quantum-}\\\textbf{pretrained 3}} \\
\midrule
GBA & 1:1 & 0.55 $\pm$ 0.02 & 0.76 $\pm$ 0.06 & 0.79 $\pm$ 0.01 & 1:6 & 0.59 $\pm$ 0.10 & 0.75 $\pm$ 0.00 & \textbf{0.79 $\pm$ 0.01} \\
KAT2A & & 0.53 $\pm$ 0.02 & 0.66 $\pm$ 0.09 & \textbf{0.69 $\pm$ 0.06} & & 0.63 $\pm$ 0.01 & 0.63 $\pm$ 0.44 & \textbf{0.66 $\pm$ 0.02} \\
ESR1 antagonist & & \textbf{0.80 $\pm$ 0.02} & 0.72 $\pm$ 0.04 & 0.79 $\pm$ 0.03 & & \textbf{0.80 $\pm$ 0.01} & 0.72 $\pm$ 0.04 & 0.77 $\pm$ 0.01 \\
MAPK1 & & 0.57 $\pm$ 0.01 & 0.67 $\pm$ 0.01 & 0.61 $\pm$ 0.01 & & 0.53 $\pm$ 0.01 & 0.65 $\pm$ 0.01 & \textbf{0.67 $\pm$ 0.03} \\
FEN1 & & \textbf{0.62 $\pm$ 0.01} & 0.51 $\pm$ 0.02 & 0.59 $\pm$ 0.04 & & 0.53 $\pm$ 0.01 & 0.60 $\pm$ 0.04 & 0.51 $\pm$ 0.03 \\
PKM2 & & 0.51 $\pm$ 0.02 & 0.51 $\pm$ 0.03 & 0.56 $\pm$ 0.07 & & 0.65 $\pm$ 0.01 & 0.56 $\pm$ 0.01 & 0.58 $\pm$ 0.02 \\
VDR & & \textbf{0.67 $\pm$ 0.02} & 0.60 $\pm$ 0.05 & 0.59 $\pm$ 0.04 & & 0.57 $\pm$ 0.02 & \textbf{0.70 $\pm$ 0.03} & 0.58 $\pm$ 0.03 \\
ALDH1 & & 0.54 $\pm$ 0.03 & 0.59 $\pm$ 0.03 & 0.52 $\pm$ 0.02 & & 0.52 $\pm$ 0.02 & \textbf{0.59 $\pm$ 0.01} & 0.56 $\pm$ 0.01 \\
\bottomrule
\end{tabularx}
\end{table*}

\begin{table*}[htbp]
\caption{\label{tab:table10} Classification performances of quantum-pretrained classical embedding methods (trained classical neural networks from NQE with the XYZ feature map). Quantum-pretrained performance outperforming both NQE and RBF is highlighted in bold.}
\scriptsize
\centering
\begin{tabularx}{\textwidth}{lcccccccc}
\toprule
\textbf{Target} &
\textbf{Class ratio} &
\makecell{\textbf{Quantum-}\\\textbf{pretrained 1}} &
\makecell{\textbf{Quantum-}\\\textbf{pretrained 2}} &
\makecell{\textbf{Quantum-}\\\textbf{pretrained 3}} &
\textbf{Class ratio} &
\makecell{\textbf{Quantum-}\\\textbf{pretrained 1}} &
\makecell{\textbf{Quantum-}\\\textbf{pretrained 2}} &
\makecell{\textbf{Quantum-}\\\textbf{pretrained 3}} \\
\midrule
GBA & 1:1 & \textbf{0.65 $\pm$ 0.01} & 0.51 $\pm$ 0.01 & 0.51 $\pm$ 0.01 & 1:6 & 0.51 $\pm$ 0.01 & 0.61 $\pm$ 0.04 & 0.61 $\pm$ 0.01 \\
KAT2A & & 0.53 $\pm$ 0.03 & \textbf{0.60 $\pm$ 0.00} & 0.51 $\pm$ 0.02 & & 0.60 $\pm$ 0.01 & \textbf{0.65 $\pm$ 0.02} & 0.58 $\pm$ 0.02 \\
ESR1 antagonist & & 0.72 $\pm$ 0.01 & 0.68 $\pm$ 0.01 & 0.75 $\pm$ 0.03 & & 0.71 $\pm$ 0.03 & \textbf{0.75 $\pm$ 0.02} & 0.73 $\pm$ 0.06 \\
MAPK1 & & \textbf{0.61 $\pm$ 0.52} & 0.54 $\pm$ 0.05 & 0.60 $\pm$ 0.04 & & 0.58 $\pm$ 0.00 & 0.57 $\pm$ 0.03 & 0.56 $\pm$ 0.02 \\
FEN1 & & 0.52 $\pm$ 0.07 & 0.57 $\pm$ 0.00 & 0.54 $\pm$ 0.01 & & 0.52 $\pm$ 0.00 & 0.57 $\pm$ 0.04 & 0.58 $\pm$ 0.02 \\
PKM2 & & 0.58 $\pm$ 0.00 & 0.58 $\pm$ 0.03 & \textbf{0.63 $\pm$ 0.00} & & \textbf{0.73 $\pm$ 0.01} & 0.60 $\pm$ 0.04 & 0.57 $\pm$ 0.02 \\
VDR & & \textbf{0.69 $\pm$ 0.01} & 0.51 $\pm$ 0.01 & 0.56 $\pm$ 0.00 & & \textbf{0.69 $\pm$ 0.00} & 0.60 $\pm$ 0.04 & 0.55 $\pm$ 0.00 \\
ALDH1 & & 0.58 $\pm$ 0.00 & 0.58 $\pm$ 0.01 & 0.53 $\pm$ 0.09 & & 0.55 $\pm$ 0.00 & 0.55 $\pm$ 0.02 & 0.55 $\pm$ 0.03 \\
\bottomrule
\end{tabularx}
\end{table*}

\subsection{Quantum kernel-based SVM results for the COVID-19 dataset}

Unlike the aforementioned analyses for the LIT-PCBA dataset, we applied SVM with the ZZ feature map and the XYZ feature map–based kernel to the COVID-19 dataset. As summarized in Table \ref{tab:table11}, SVM with the ZZ feature map kernel achieved balanced accuracies of 0.71 $\pm$ 0.00 and 0.76 $\pm$ 0.00 for the 4-qubit and 8-qubit models, respectively. When employing the PQK with the ZZ feature map, the performance increased to 0.83 $\pm$ 0.00 for the 4-qubit configuration and 0.80 $\pm$ 0.00 for the 8-qubit configuration. In contrast, SVM using the XYZ feature map kernel without the PQK showed accuracies of 0.50 $\pm$ 0.00 and 0.50 $\pm$ 0.00, which were close to the random chance level. The PQK-augmented XYZ feature map improved the classification accuracy to 0.72 $\pm$ 0.00 and 0.73 $\pm$ 0.00 for the 4-qubit and 8-qubit cases, respectively. For comparison, classical SVM models showed lower overall performances, with balanced accuracies of 0.59 $\pm$ 0.00 (linear kernel), 0.60 $\pm$ 0.00 (RBF kernel with PCA), 0.62 $\pm$ 0.00 (linear kernel without PCA), and 0.65 $\pm$ 0.00 (RBF kernel without PCA). These results indicate that the quantum embedding approaches, particularly those employing the PQK, consistently outperformed the classical counterparts in this dataset.

\begin{table*}[htbp]
\caption{\label{tab:table11} Classification performances of SVM with quantum kernels evaluated on the COVID-19 dataset.}
\scriptsize
\centering
\begin{tabular}{lc}
\toprule
\textbf{Condition} & \textbf{\makecell{Balanced accuracy\\(4-qubits  / 8-qubits)}}\\
\midrule
PCA + SVM (ZZ feature map kernel) & 0.71 $\pm$ 0.00 / 0.76 $\pm$ 0.00 \\
PCA + SVM (PQK\footnotemark[1] with the ZZ feature map) & 0.83 $\pm$ 0.00 / 0.80 $\pm$ 0.00 \\
PCA + SVM (XYZ feature map kernel) & 0.50 $\pm$ 0.00 / 0.50 $\pm$ 0.00 \\
PCA + SVM (PQK with the XYZ feature map) & 0.72 $\pm$ 0.00 / 0.73 $\pm$ 0.00 \\
\midrule
PCA + SVM (linear kernel) & 0.59 $\pm$ 0.00\\
PCA + SVM (RBF kernel) & 0.60 $\pm$ 0.00\\
SVM (linear kernel) & 0.62 $\pm$ 0.00\\
SVM (RBF kernel) & 0.65 $\pm$ 0.00\\
\bottomrule
\end{tabular}
\footnotetext[1]{PQK : the projected quantum kernel}
\end{table*}

\section{\label{sec:level3}Discussion}

\subsection{Comparisons between the quantum and classical embedding approach}

To provide the rationale for applying the quantum embedding method to the ligand-based virtual screening problem with the LIT-PCBA dataset, we first compared the embedding performance between the quantum embedding method, which utilizes NQE, and its classical counterpart, which employs RBF kernel with the classical neural network. Since the same classical neural network architecture was used in both NQE and RBF kernel, the key difference between the two approaches lies solely in their embedding methods, that is, the ZZ feature map and XYZ feature map for the quantum approach versus RBF kernel for the classical approach. Based on this difference, the classification performance of the binary classifiers was compared as an evaluation of embedding performance (the 8-qubit QCNN for the quantum model and the single-layer network for the classical model). 

Among the 32 comparison conditions (1:1 and 1:6 class ratio conditions for each biological target and two types of quantum feature maps), the QCNN with NQE achieved higher classification performance than the single-layer network with RBF kernel in most conditions, except for the 1:1 ratio case with the XYZ feature map. While the XYZ feature map-based NQE under 1:1 ratio showed lower performance than its classical counterpart for all biological targets, the overall trend indicates that quantum embedding methods utilizing classically intractable quantum feature maps generally outperform classical methods across various feature maps and class ratios. In addition, for the COVID-19 dataset analyzed using SVM with the 4- and 8-qubits ZZ feature map kernel, the XYZ feature map kernel, and the PQK showed improved classification performances compared to SVMs using RBF and linear kernels. Similar to the model performance with the LIT-PCBA dataset, these performance trends suggest that the QSVM can achieve enhanced performance, particularly under limited and imbalanced data conditions, which are commonly encountered in drug discovery tasks. 

These findings are consistent with previous studies suggesting that the power of quantum kernels arises from classically intractable feature mappings implemented by quantum circuits \cite{schuld2019quantum, hubregtsen2022training}. Such mappings enable high-dimensional representations in quantum Hilbert space, improving the separability and classification of molecular features. Based on this interpretation, our results provide empirical evidence supporting the effectiveness of the quantum kernel in ligand-based virtual screening tasks.

\subsection{Hybrid quantum–classical approaches for enhanced data embedding}

From the classical neural networks trained by NQE with the LIT-PCBA dataset, we proposed and evaluated three hybrid embedding approaches. The design of these approaches was inspired by the concept of transfer learning, which is widely applied in classical machine learning to reduce computational costs by reusing pretrained models rather than training from scratch, as suggested in previous studies \cite{kim2023classical, mari2020transfer}. These studies demonstrated that transferring trained parameters between classical and quantum models can be effective when applied to toy datasets. Building on this motivation, we extended this paradigm to the development of hybrid embedding strategies for ligand-based drug discovery.

In the first quantum-pretrained approach (quantum-pretrained 1), the pretrained neural network from NQE was fine-tuned under RBF with the classical neural network scheme, and a single-layer network was appended for binary classification. This approach assumed that the optimized weight parameters obtained from quantum fidelity optimization could enhance model performance when fine-tuned with the classical RBF kernel. Among 24 experimental conditions (eight biological targets, two class ratios, and two quantum feature maps), the quantum-pretrained 1 model showed higher classification performance than both the quantum and classical embedding methods in nine cases. This indicates that transferring and fine-tuning quantum-trained parameters within a classical kernel framework can effectively improve embedding quality. In the second (quantum-pretrained 2) and third (quantum-pretrained 3) approaches, the pretrained neural network from NQE was incorporated into a single-layer network, with fixed parameters in quantum-pretrained 2 and trainable parameters in quantum-pretrained 3. These two approaches were designed to examine how the trained neural network in the quantum model influences classical classification performance. Under quantum-pretrained 2, improvements were observed in five conditions, and quantum-pretrained 3 achieved similar enhancements in another 5 conditions. These findings imply that the fine-tuning approach allows the data representation acquired during quantum embedding optimization to be effectively transferred and utilized within classical frameworks, thereby improving model generalization and representation efficiency.

\subsection{Potential of the proposed embedding framework for the ligand-based drug discovery}
From a practical standpoint, the proposed quantum and hybrid embedding frameworks show several advantages that align closely with the requirements of ligand-based virtual screening. First, the classically intractable quantum feature maps, such as the ZZ feature map and the XYZ feature map, provide richer molecular embeddings that can capture complex physicochemical patterns beyond what classical kernels can represent. Second, the robustness of the quantum classifiers under class-imbalanced conditions indicates that these models may offer practical benefits in real-world datasets, where inactive compounds are typically overrepresented. Finally, the hybrid transfer-learning schemes further extend the applicability of the framework by enabling the reuse of pretrained neural embeddings across quantum and classical models. This allows efficient adaptation of the trained embeddings to new ligand datasets without the need for complete retraining, reducing the computational overhead typically associated with large-scale virtual screening. Collectively, these advantages suggest that the proposed quantum and hybrid approaches hold promise for integration into future drug discovery pipelines as efficient and expressive embedding modules, thereby bridging quantum-enhanced representation learning with classical screening infrastructures.

\section{\label{sec:level4}Conclusion}
In this study, we investigated a family of quantum and hybrid data-embedding strategies for improving ligand-based virtual screening. Our results show that trainable quantum embeddings derived from NQE consistently outperform classical RBF-based embeddings across multiple biological targets, demonstrating the representational advantages offered by parameterized quantum feature maps. We further introduced quantum-pretrained classical embeddings, which reuse NQE-trained neural networks within purely classical models, and found that these transfer-learning variants often provide additional gains in limited-data and class-imbalanced settings.

Taken together, these findings indicate that both quantum and quantum-pretrained embeddings can serve as effective and data-efficient molecular representation tools, offering practical value for early-stage drug discovery. An important direction for future work is to better understand the relationship between quantum embedding geometry and underlying molecular structure, thereby enabling more interpretable and task-aligned quantum feature maps. Experimental validation on quantum hardware and the development of hardware-aware embedding designs will further advance the applicability of these methods in real-world screening pipelines.

\begin{acknowledgments}
This research was supported by the education and training program of the Quantum Information Research Support Center, funded through the National research foundation of Korea (NRF) by the Ministry of science and ICT (MSIT) of the Korean government (RS-2023-NR057243), by Institute of Information \& communications Technology Planning \& evaluation (IITP) grant funded by the Korea government (No. 2019-0-00003, Research and Development of Core Technologies for Programming, Running, Implementing and Validating of Fault-Tolerant Quantum Computing System), the National Research Foundation of Korea (RS-2025-02309510), the Ministry of Trade, Industry, and Energy (MOTIE), Korea, under the Industrial Innovation Infrastructure Development Project (RS-2024-00466693), and by Korean ARPA-H Project through the Korea Health Industry Development Institute (KHIDI), funded by the Ministry of Health \& Welfare, Korea (RS-2025-25456722).
\end{acknowledgments}

\section*{Data Availability}
The LIT-PCBA and COVID-19 dataset used in this study are publicly available open-source datasets. Details of the datasets and access information are provided in the corresponding references. 

\section*{Code Availability}
The source code for this project is publicly available on the \url{https://github.com/Jungguchoi/Optimizing-Quantum-Data-Embeddings-for-Ligand-Based-Virtual-Screening}

\bibliography{reference}

\appendix

\clearpage
\renewcommand\appendixname{}
\renewcommand{\thesection}{Appendix A}
\renewcommand{\thefigure}{A.\arabic{figure}} 
\renewcommand{\thetable}{A\arabic{table}} 
\setcounter{figure}{0}
\setcounter{table}{0}

\begin{widetext}
\section{Molecular features and trace distance results from the quantum embedding}
\label{appendix:appendixA}

\begin{table*}[htbp]
\renewcommand{\arraystretch}{1.5}
\caption{List of molecular features calculated from the LIT-PCBA dataset}
\begin{ruledtabular}
\begin{tabular}{llll}
\textbf{No.} & \textbf{Feature} & \textbf{No.} & \textbf{Feature} \\
\colrule
1  & Num\_C & 21 & MolLogP \\
2  & Num\_N & 22 & MolWt \\
3  & Num\_O & 23 & FpDensityMorgan1 \\
4  & Num\_P & 24 & FpDensityMorgan2 \\
5  & Num\_S & 25 & FpDensityMorgan3 \\
6  & Num\_F & 26 & MaxAbsPartialCharge \\
7  & Num\_Cl & 27 & MinAbsPartialCharge \\
8  & Num\_Br & 28 & NumValenceElectrons \\
9  & Num\_I & 29 & BertzCT \\
10 & Single\_Bonds & 30 & BalabanJ \\
11 & Double\_Bonds & 31 & Chi0 \\
12 & NumStereoE & 32 & Chi1 \\
13 & Num\_Aromatic\_Atoms & 33 & Chi2n \\
14 & Aromatic\_Proportion & 34 & Chi3n \\
15 & NumRotatableBonds & 35 & HallKierAlpha \\
16 & Total\_NH\_OH & 36 & Ipc \\
17 & Total\_N\_O & 37 & Kappa1 \\
18 & NumHydrogenAcceptors & 38 & Kappa2 \\
19 & NumHydrogenDonors & 39 & Kappa3 \\
20 & NumofHeteroatoms &  &  \\
\end{tabular}
\end{ruledtabular}
\label{tab:appendixA1}
\end{table*}
\clearpage

\begin{table*}[htbp] 
\renewcommand{\arraystretch}{1.9}
\caption{Trace distance comparisons from the NQE with the ZZ feature map evaluated on the LIT-PCBA dataset.} 
\begin{ruledtabular} 
\scriptsize
\begin{tabular}{l l l l l l l l}
\textbf{Class ratio} & \textbf{Target} & \makecell{\textbf{Before training} \\ \textbf{(train/test)}}\footnotemark[1] & \makecell{\textbf{After training} \\ \textbf{(train/test)}} & \textbf{Class ratio} & \textbf{Target} & \makecell{\textbf{Before training} \\ \textbf{(train/test)}} & \makecell{\textbf{After training} \\ \textbf{(train/test)}} \\
\colrule 1:1 & GBA & 0.0004 / 0.0005 & 0.5444 / 0.5535 & 1:6 & GBA & 0.0001 / 0.0001 & 0.1962 / 0.2339 
\\ & KAT2A & 0.0003 / 0.0004 & 0.2376 / 0.4714 & & KAT2A & 0.0002 / 0.0003 & 0.1782 / 0.2584 
\\ & ESR1 antagonist & 0.0007 / 0.0006 & 0.5080 / 0.3435 & & ESR1 antagonist & 0.0004 / 0.0007 & 0.2970 / 0.3450 
\\ & MAPK1 & 0.0004 / 0.0004 & 0.2939 / 0.2873 & & MAPK1 & 0.0004 / 0.0003 & 0.2992 / 0.5676 
\\ & FEN1 & 0.0003 / 0.0003 & 0.1206 / 0.1010 & & FEN1 & 0.0002 / 0.0003 & 0.3972 / 0.4019 
\\ & PKM2 & 0.0004 / 0.0005 & 0.3950 / 0.4456 & & PKM2 & 0.0003 / 0.0004 & 0.2154 / 0.2535 
\\ & VDR & 0.0003 / 0.0002 & 0.2443 / 0.4515 & & VDR & 0.0002 / 0.0003 & 0.2442 / 0.2867 \\ & ALDH1 & 0.0001 / 0.0002 & 0.1728 / 0.1170 & & ALDH1 & 0.0001 / 0.0001 & 0.0678 / 0.0790 \\ 
\end{tabular} 
\end{ruledtabular} 
\footnotetext[1]{(train/test) : the training and test dataset}
\label{tab:appendixA2} 
\end{table*}

\begin{table*}[htbp]
\renewcommand{\arraystretch}{1.9}
\caption{Trace distance comparisons from the NQE with the XYZ feature map evaluated on the LIT-PCBA dataset.} 
\begin{ruledtabular} 
\scriptsize
\begin{tabular}{l l l l l l l l} 
\textbf{Class ratio} & \textbf{Target} & \makecell{\textbf{Before training} \\ \textbf{(train/test)}}\footnotemark[1] & \makecell{\textbf{After training} \\ \textbf{(train/test)}} & \textbf{Class ratio} & \textbf{Target} & \makecell{\textbf{Before training} \\ \textbf{(train/test)}} & \makecell{\textbf{After training} \\ \textbf{(train/test)}} \\
\colrule 1:1 & GBA & 0.0004 / 0.0009 & 0.0033 / 0.0072 & 1:6 & GBA & 0.0007 / 0.0009 & 0.3199 / 0.4161 
\\ & KAT2A & 2.1e-05 / 3.6e-05 & 0.0013 / 0.0015 & & KAT2A & 5.8e-05 / 7.1e-05 & 0.1311 / 0.1364 
\\ & ESR1 antagonist & 0.0038 / 0.0038 & 0.0098 / 0.0074 & & ESR1 antagonist & 0.0029 / 0.0025 & 0.2316 / 0.2879 
\\ & MAPK1 & 0.0012 / 0.0012 & 0.0012 / 0.0017 & & MAPK1 & 0.0010 / 0.0012 & 0.1446 / 0.1407 
\\ & FEN1 & 0.0003 / 0.0004 & 0.0006 / 0.0007 & & FEN1 & 1.0e-05 / 1.5e-05 & 0.4132 / 0.3987 
\\ & PKM2 & 0.0005 / 0.0005 & 0.0011 / 0.0012 & & PKM2 & 2.0e-05 / 3.0e-05 & 0.0909 / 0.1137 
\\ & VDR & 0.0003 / 0.0004 & 0.0036 / 0.0041 & & VDR & 1.0e-05 / 9.4e-05 & 0.2014 / 0.1940 \\ & ALDH1 & 1.9e-05 / 2.0e-05 & 0.0009 / 0.0009 & & ALDH1 & 3.0e-05 / 5.0e-05 & 0.0447 / 0.0516 \\ 
\end{tabular} 
\end{ruledtabular} 
\footnotetext[1]{(train/test) : the training and test dataset}
\label{tab:appendixA3} 
\end{table*}

\renewcommand\appendixname{}
\renewcommand{\thesection}{Appendix B}

\section{Applied hyperparameters and optimized structures of the classical neural network for the quantum, classical, and quantum-pretrained classical embedding method}
\label{appendix:appendixB}

Appendix B is provided as supplementary materials and are available at the following GitHub repository:
\url{https://github.com/Jungguchoi/Optimizing-Quantum-Data-Embeddings-for-Ligand-Based-Virtual-Screening/blob/main/AppendixB.pdf}

\renewcommand\appendixname{}
\renewcommand{\thesection}{Appendix C}

\section{Training and validation loss graphs for the embedding method and the evaluation model}
\label{appendix:appendixC}

Appendix C is also provided as supplementary materials and are available at the following GitHub repository:
\url{https://github.com/Jungguchoi/Optimizing-Quantum-Data-Embeddings-for-Ligand-Based-Virtual-Screening/blob/main/AppendixC.pdf}

\end{widetext}
\clearpage
\nocite{*}

\end{document}